\documentclass[12pt,epsf]{article}
\newcommand{\TT}{{\cal T}}

\newcommand{\be}{\begin{equation}}
\newcommand{\ee}{\end{equation}}
\newcommand{\ben}{\begin{eqnarray}\displaystyle}
\newcommand{\een}{\end{eqnarray}}
\newcommand{\refb}[1]{(\ref{#1})}
\newcommand{\p}{\partial}

\newcommand{\s}{\sigma}

\newcommand{\Tr}{\hbox{Tr}}
\newcommand{\DT}{\Delta T}
\newcommand{\diag}{\hbox{diag}}

\begin{document}

{}~ \hfill\vbox{\hbox{hep-th/0203108}
\hbox{UUITP-04/02}}\break

\vskip 1.cm

\centerline{\large \bf Stretched strings in tachyon condensation models}
\vspace*{1.5ex}

\vspace*{4.0ex}

\centerline{\large \rm Joseph A. Minahan\footnote{E-mail:
joseph.minahan@teorfys.uu.se}}
\vspace*{2.5ex}
\centerline{\large \it Department of Theoretical Physics}
\centerline{\large \it Box 803, SE-751 08 Uppsala, Sweden}
\vspace*{3.0ex}
\medskip
\centerline {\bf Abstract}

In this note, we consider the two derivative truncation of boundary
string
field theory for the unstable D9 brane in Type IIA string theory.  We
construct
multiples of the stable codimension 1 solitons that correspond to
stacks of D8
branes.  We find the fluctuation modes that correspond to open strings
stretching between the branes, and find that their masses are
consistent
with the string tension.  We show that these modes are localized
halfway between the branes and that their width is independent of the
brane separation.

\bigskip

\vfill \eject
\baselineskip=17pt

%\sectiono{Introduction}

Simple field theory models have proven to be quite useful in giving
qualitative and surprisingly  good quantitative descriptions for tachyon
condensation\footnote{For early pioneering work in
the field of tachyon condensation see \cite{pioneer}.}
 in open string field theory 
\cite{0002117}-\cite{0112088}.  These models turn out to be
two derivative truncations of boundary string field theory
\cite{0009103}-\cite{bsftother}. In particular, the
model described in \cite{0009246} is the two derivative
truncation of boundary superstring field theory \cite{0010108}.
The lagrangian for  this model is given by
\be
-\TT_9 \left(\frac{1}{2}\p_\mu T\p_\mu T+1\right)e^{-T^2/4}
\ee
where $\TT_9$ is the
tension
of the the unstable D9 brane.  The potential has an unstable vacuum at
$T=0$,
 where the tachyon mass squared
is $-1/2$ (we have set $\alpha'=1$).  It also has stable
vacuua at $T=\pm\infty$, where the mass of the tachyon becomes infinite.  
There is a stable codimension one soliton solution with
a tension $\TT_8=2\sqrt{2\pi}\TT_9$, a result reasonably close to the 
actual value of $\TT_8=\sqrt{2}\pi\TT_9$.
It was
also shown in \cite {0009246} that the fluctuation modes about this
soliton have integer 
mass squared level spacing.  Thus, this model
nicely conforms with the Sen conjectures \cite{senconj}.

Of course, open string field theory should contain an infinite number of
fields.  In \cite{0011226} a prescription was given for including higher level
fields into the models.  For instance, it was argued that including gauge
fields with the Lagrangian
\be\label{gaugel}
-\frac{1}{4}F_{\mu\nu}F^{\mu\nu}e^{-T^2/4}
\ee
led to fluctuation modes on the codimension one brane with the correct
mass squared spectra.  Moreover, it was shown that a zero momentum mode
for the gauge field was gauged away on the D8 brane
but was compensated by the zero mode coming from the tachyon
field.  Some preliminary evidence was also given 
for how to include higher level fields in the model.  In any event, the
surprising conclusion is reached that dropping the higher derivative
terms does
not seem to drastically effect the physics, at least at tree-level.  Dropping
the higher derivatives changes the tensions of the branes and fattens them out,
but it does not seem to effect the spectrum.

The ``open string'' modes  that we find in these models are localized around
the brane, that is, the center of the soliton.  One interesting question is
what do the stretched strings look like in these models.  By stretched
string we mean
a mode that corresponds to a string stretching between two D-branes separated
by some distance.  The naive guess is that the mode is more or less evenly 
distributed between the branes.  This turns out not to be the case.  Instead
we will show that the mode {\it is } localized halfway between the branes,
with the localization width independent of the brane separation.  This last
fact seems to indicate that the modes will be localized, not stretched,
even when higher 
derivative terms are taken into account.  

Nevertheless, the modes get an extra contribution to their
mass-squared which is precisely the amount expected for a string
stretched between two D-branes.

%In order to have multiple branes, the tachyon field needs to be extended to
%an $N\times N$ tensor that transforms under the adjoint of a $U(N)$ gauge
%group.  As we will see, ordering ambiguities will arise.  However,  we can 
%remove some of the ambiguity by insisting on a reasonable spectrum for the
%fluctuation modes.

%\sectiono{Multiple branes and their fluctuations}

%In this section we consider the problem of multiple branes in both the bosonic and  the supersymmetric truncated open string field theories.

%\subsection{Multiple branes }

In order to describe multiple brane solutions, the tachyon field needs to
be generalized to an $N$ by $N$ tensor field
\cite{9812135}-\cite{0110092},\cite{0010108}.  
The tachyon potential is
then \cite{0010108}
\be\label{tenspot}
V(T)=\Tr(e^{-T^2/4}).
\ee
Generalizing the kinetic term requires some care because of ordering ambiguities.  One consideration is to find a solvable spectrum.  We will see that
\be\label{tenskin}
\frac{1}{2}\Tr(\p_\mu T e^{-T^2/8}\p_\mu T e^{-T^2/8})
\ee
serves this purpose.

Nontrivial codimension 1 
solutions to the equations of motion are the diagonal matrices
\be
T=\diag(T_1,T_2...T_N)
\ee
where 
\be\label{tachsol}
T_i=\pm\sqrt{2}(x - x_i)
\ee
and where $x_i$ is the location of the $i^{\rm th}$ brane.  
The $+$ ($-$) sign corresponds to a D8 brane (anti-brane).

Let us begin by assuming that $N=2$, our results
being easily generalized to higher $N$.
In this case, we can rewrite $T$ as
\be
T=T_0\s_0+\DT \s_3
\ee
where
\be
\s_0=\left(\begin{array}{cc}1&0\\0&1\end{array}\right)\qquad\qquad
\s_3=\left(\begin{array}{cc}1&0\\0&-1\end{array}\right)\qquad\qquad.
\ee

We now look for fluctuations about these solutions.  The diagonal fluctuations
are exactly as in \cite{0009246,0011226}.  However, the off-diagonal fluctuations are 
nontrivial if $\DT\ne0$.  In order to find these, let us write
the off-diagonal fluctuation piece as
\be
T_{fl}=\left(\begin{array}{cc}0&T_+\\T_-&0\end{array}\right).
\ee
We now make use of the fact
that 
\be\label{niceprop}
\Tr\left(\left(\begin{array}{cc}0&A\\B&0\end{array}\right)(\s_3)^n\left(\begin{array}{cc}0&A\\B&0\end{array}\right)(\s_3)^m\right)
=AB\left[(-1)^n+(-1)^m\right],
\ee
which among other things implies that
\begin{eqnarray}\label{niceprop2}
&&\Tr\Bigg[\left(\begin{array}{cc}0&A\\B&0\end{array}\right)e^{-T_0^2/8-\DT^2/8-T_0\DT\s_3/4}
\left(\begin{array}{cc}0&A\\B&0\end{array}\right)e^{-T_0^2/8-\DT^2/8-T_0\DT\s_3/4}\Bigg]\nonumber\\
&&\qquad\qquad
=
2AB\exp(-T_0^2/4-\DT^2/4).
\end{eqnarray}
To quadratic order in the off-diagonal fluctuations, the lagrangian is
\be\label{offdf}
-\left((\p_\mu T_{+}\p_\mu T_{-}+\frac{1}{4}[(\p_\mu T_0)^2-(\p_\mu \DT)^2-2]T_{+}T_{-}\right)e^{-T_0^2/4-\DT^2/4}
\ee
where we have used the equations of motion for $T_0$ and
$\DT$,
\begin{eqnarray}\label{eomt}
\p^2 T_0-\frac{1}{4}T_0\left[(\p T_0)^2+(\p
\DT)^2\right]-\frac{1}{2}\DT\p_\mu T_0\p_\mu\DT +\frac{1}{2}T_0&=&0
\nonumber\\
\p^2 \DT-\frac{1}{4}\DT\left[(\p T_0)^2+(\p
\DT)^2\right]-\frac{1}{2}T_0\p_\mu T_0\p_\mu\DT +\frac{1}{2}\DT&=&0.
\end{eqnarray}
Making further use of the equations of motion and
integrating by parts, we
find that the expression in \refb{offdf} can be rewritten as
\be\label{offdfl}
-\p_\mu\left(\frac{T_{+}}{\DT}\right)\p_\mu \left(\frac{T_{-}}{\DT}\right)e^{-T_0^2/4-\DT^2/4}(\DT)^2.
\ee

The $U(N)$ symmetry of the tachyon field is actually a gauge symmetry,
so accordingly, the ordinary derivatives should be replaced by covariant
derivatives.  We will assume that the kinetic term has the form
\be\label{tensking}
-\frac{1}{2}\Tr([\p_\mu-iA_\mu, T] e^{-T^2/8}[\p_\mu-iA_\mu, T]e^{-T^2/8}).
\ee
We then expect {\it all} 
of the off-diagonal fluctuations of $T$ to be eaten
by the gauge fields.  Again, let us consider the case that $N=2$
and let us find the terms quadratic in $A_\mu$ and the tachyon fluctuations.
Using the properties in \refb{niceprop} and \refb{niceprop2}, we find
the quadratic off-diagonal piece to be
\be\label{offdiag}
-4\left(A_{+\mu}+\frac{i}{2}\p_\mu\left(\frac{T_{+}}{\DT}\right)\right)
\left(A_{-\mu}-\frac{i}{2}\p_\mu\left(\frac{T_{-}}{\DT}\right)\right)
e^{-T_0^2/4-\DT^2/4}(\DT)^2.
\ee

To show the importance of ordering in the tachyon kinetic terms, let
us suppose that the appropriate term in the lagrangian were
\be 
\frac{1}{2}\Tr([\p_\mu-iA_\mu, T]^2 e^{-T^2/4}]).
\ee
In this case, one would find that the off-diagonal fluctuations have
the form 
\be
4\left(A_{+\mu}+\frac{i}{2}\p_\mu\left(\frac{T_{+}}{\DT}\right)\right)
\left(A_{-\mu}-\frac{i}{2}\p_\mu\left(\frac{T_{-}}{\DT}\right)\right)
e^{-T_0^2/4-\DT^2/4}\cosh(T_0\DT/4)\DT^2,
\ee
and so would clearly have a different spectrum than that coming from
\refb{offdiag}.

Now consider the kinetic term for the gauge fields.  Again, there will be
ordering ambiguities.  There is also some ambiguity as to how even the
diagonal action should be chosen.  For the D9 brane, one expects the
quadratic piece of the lagrangian to come from expanding the Born-Infeld
lagrangian
\be\label{BIaction}
\sqrt{\det(\eta_{\mu\nu}-2\pi
F_{\mu\nu})}\left(\frac{1}{2}G^{\mu\nu}\p_\mu Te^{-T^2/8}\p_\nu T
e^{-T^2/8}+
e^{-T^2/4}\right),
\ee
where $G^{\mu\nu}$ is the open string metric which is the symmetric
part
of \cite{9908142} 
\be
\left(\eta_{\mu\nu}-2\pi
F_{\mu\nu}\right)^{-1}. 
\ee
 The problem with this
formulation is that the D8 solution for $T$ will have no quadratic
piece for the gauge fields; the contribution from the kinetic term
cancels off the contribution from the potential.  These difficulties are probably due to
our two derivative limitation as well as the assumption in
\refb{BIaction} that the field strengths are constant.

We will instead assume that the quadratic gauge lagrangian has the
form
\be\label{gaugekin}
-\frac{(2\pi)^2}{4}\Tr
\left(\frac{1}{2}F_{\mu\nu}\p_\gamma T e^{-T^2/8}\p_\gamma T F_{\mu\nu}e^{-T^2/8}+F_{\mu\nu}e^{-T^2/8}F_{\mu\nu}e^{-T^2/8}\right).
\ee
This form of the lagrangian is consistent with reduction from the unstable
D9 to the stable D8 branes.
The ordering in \refb{gaugekin} leads to a particularly simple form
for
the off-diagonal gauge fluctuations,
\be
(2\pi)^2\left(\p_\mu A_{+\nu}\p_\nu A_{-\nu}-\p_\mu
A_{+\nu}\p_\nu
A_{-\mu}\right)e^{-T_0^2/4-\DT^2/4}\left(\frac{1}{2}\p_\mu T_0\p_\mu
T_0+\frac{1}{2}\p_\mu\DT\p_\mu\DT+1\right).
\ee
Defining $A'_{+\mu}$ as
\be
A'_{+\mu}=A_{+\mu}+\frac{i}{2}\p_\mu\left(\frac{T_{+}}{\DT}\right),
\ee
the complete quadratic lagrangian for the off-diagonal gauge
fluctuations is
\begin{eqnarray}\label{quadfl}
&&-\Bigg[(2\pi)^2\left(\p_\mu A'_{+\nu}\p_\mu A'_{-\nu}-\p_\mu
A'_{+\nu}\p_\nu
A'_{-\mu}\right)\left(\frac{1}{2}(\p T_0)^2
+\frac{1}{2}(\p\DT)^2+1\right)
\nonumber\\
&&\qquad\qquad\qquad\qquad\qquad
+4A'_{+\mu}A'_{-\mu}\DT^2\Bigg]e^{-T^2_0/4-\DT^2/4}.
\end{eqnarray}

We can now use \refb{quadfl} to compute the spectrum of $A'_{\pm\mu}$.
Let us take the classical solutions in \refb{tachsol}, 
and assume that we have two D8 branes, one at $x=x_0/2$ and the other
at
$x=-x_0/2$.  Thus,
\begin{eqnarray}
T_1&=&\sqrt{2}(x-x_0/2)
\nonumber\\
T_2&=&\sqrt{2}(x+x_0/2),
\end{eqnarray}
and so
\begin{eqnarray}
T_0&=&\sqrt{2}x
\nonumber\\
\DT&=&-\frac{\sqrt{2}}{2}x_0.
\end{eqnarray}
The equations of motion for $T_{+}$ lead to the equation
\be
\p_\mu (A'_{+\mu}e^{-x^2/2})=0.
\ee
Substituting this into the equation of motion for $A'_{+\nu}$, we
have
\be\label{A+eom}
-\p^2A'_{+\nu}+x\p_x
A'_{+\nu}+A'_{+,x}\delta_{\nu x}
+\frac{x_0^2}{(2\pi)^2}A'_{+\nu}=0.
\ee
Letting $A_{+\nu}=B_\nu e^{-x^2/4}$ we
find the equation
\be
-\p^2 B_\nu
+\left[\frac{1}{4}x^2-\frac{1}{2}+\delta_{\nu x}+\frac{x_0^2}{(2\pi)^2}\right]B_\nu=0.
\ee
Hence the mass spectra for the modes is
\be
m^2=n+\frac{x_0^2}{(2\pi)^2}\qquad\qquad n\ge 0
\ee
for polarizations along the brane ($\nu\ne x$)
and 
\be
m^2=n+1+\frac{x_0^2}{(2\pi)^2}\qquad\qquad n\ge 0
\ee
for the transverse polarization $(\nu=x)$.
Since the string tension is $1/(2\pi)$, we obtain the
desired contribution to the mass coming from a string stretching
between the two branes separated by a distance $x_0$.
Note that the massive modes of $T_+$ and $T_-$  have been eaten by the $B_x$
modes while
the massless modes of $T_+$ and $T_-$ 
have been Higgsed by the massless gauge mode.

Let us examine a few facts that might seem surprising about these
stretched
modes.  First of all, they are not really stretched.  If one were to
consider a classical wave corresponding to one of these modes, they
would find the energy density localized at $x=0$.  This   is 
half-way between the branes, which
contrasts with the diagonal modes which are localized on the
branes.   One also finds that the width of the modes is
independent of the brane
separation.

However, there is no suppression of the interactions between the charged
and the diagonal modes, even at large D-brane separation.  For
instance, consider the full nonabelian lagrangian coming from
\refb{gaugekin}.  Writing the gauge field as
\be
A_\mu=\left(\begin{array}{cc}A_{3\mu}&A_{+\mu}\\A_{-\mu}&-A_{3\mu}\end{array}\right),
\ee
the cubic interaction term coming from \refb{gaugekin} is
\be
2i(2\pi)^2A_{3\nu}\left[A_{+\mu}\p_{[\nu}
A_{-\mu]}-A_{-\mu}\p_{[\nu}A_{+\mu]}
\right]e^{-T_0^2/4-\DT^2/4}.
\ee
The kinetic term for $A_{3\mu}$ comes with a gaussian factor
localizing the mode around the branes.  But the normalized modes for
$A_{3\mu}$
have no such suppression.  For instance, the zero mode has $A_{3\mu}$
constant in the $x$ direction.  Hence the 3 point interaction is
not suppressed even though the charged modes and the neutral modes
are
localized at different positions in $x$.

It is straightforward to generalize these results to multiple branes.
In
this case, the off-diagonal fluctuations corresponding to strings
attached
to branes $i$ and $j$ are
\begin{eqnarray}\label{quadgen}
&&\Bigg[2(2\pi)^2\left(\p_\mu A'_{ij,\nu}\p_\nu A'_{ji,\nu}-\p_\mu
A'_{ij,\nu}\p_\nu
A'_{ji,\mu}\right)
\nonumber\\
&&\qquad\qquad\qquad\qquad\qquad
+A'_{ij,\nu}A'_{-\nu}(T_i-T_j)^2\Bigg]e^{-T^2_i/8-T_j^2/8}.
\end{eqnarray}
This will lead to the anticipated spectrum for the stretched strings.

One can also consider the case where the transverse direction is
compactified
on a circle such that
\be
x\equiv x+2\pi R.
\ee 
The analysis will be similar to that in \cite{9611042} for
matrix models on the circle.  Now the gauge group must be extended to
$U(\infty)$.  A stable D8 brane on the circle is given by the infinite
diagonal matrix
\be
T=\diag(..,T_n,T_{n+1}...)\qquad\qquad T_n=\sqrt{2}(x-2\pi nR).
\ee
Hence, the solution is invariant under the shift $x\to x+2\pi R$ up to
a gauge transformation.  An open string wound $n$ times around the
circle is then given by a
superposition
of the $A_{i,i+n}$ modes.  

Note that the center of mass $U(1)$ is not present in \refb{quadgen}.  The
usual
picture for an open string stretched between two branes is that of a
flux tube between opposite charges of the diagonal $U(1)$.  That is not the
picture
that emerges here.  Instead, everything operates as an adjoint 
Higgsing of the $SU(N)$ group.  Of course, this is expected 
for the low energy effective gauge theory on the world volume, 
but it is nonetheless surprising to see it
happen for the stringy modes as well.

This also raises questions about how the closed strings will appear
\cite{9901159}-\cite{0106103}.
One could certainly look for closed strings in a D brane vacuum.  For
example,
in the compact case, there should be closed strings that wrap around
the compact direction.  It seems from the analysis here that these are
more
likely to come from the $SU(\infty)$ sector and not the COM $U(1)$,
since the decay of
a wound open string into an unwound open string
and a  closed string with zero
momentum
in the transverse direction still needs to
preserve
the winding.  The wound open string modes are localized, either on the brane
or diametrically opposite from it.  One would expect a closed string to be
delocalized on the circle. It would be interesting to see how this
could occur, if indeed it can occur when only considering a finite
number of the open string fields.  One might need the full
machinery
of cubic string field theory \cite{cubiccalc} or vacuum string
field theory \cite{vsft} to see this.

It is also possible that the localization is gauge dependent.  For
instance,
in boundary string field theory, the width of a D-brane has zero size
\cite{0009148},
but
in cubic string field theory, the 
width is nonzero \cite{cubiccalc}.  These theories must be related by a gauge transformation, so
the
width of the D-brane must be a gauge artifact.  
Likewise, the widths of fluctuation modes might also be gauge
artifacts and that a gauge exists where the mode is evenly distributed
between the branes.

\bigskip
\noindent {\bf Acknowledgments}:
I would like to thank the CTP at MIT and the theory group at Harvard
 for hospitality during the course of 
this work.   I would also like to thank B. Zwiebach for many helpful
comments
on the manuscript.  This research was
supported in part by the NFR.

\end{document}